\documentclass{sf2a-conf}
\usepackage{graphicx}

\begin{document}
\TitreGlobal{SF2A 2007}
\title{Predictions of very high energy $\gamma$-ray fluxes\\
  for three Active Galactic Nuclei}


\author{J.-P. Lenain} \address{LUTH, Observatoire de Paris, CNRS, Universit{\'e} Paris Diderot; 5 Place Jules Janssen, 92190 Meudon, France\\ (jean-philippe.lenain@obspm.fr)}

\runningtitle{Predictions of VHE $\gamma$-ray fluxes for 3 AGNs}

\setcounter{page}{1}

\index{Lenain J.-P.}

\maketitle

\begin{abstract}
  M\,87 is the first extragalactic source detected at the TeV that is not a blazar. To account for the recent observations of M\,87 made by the High Energy Stereoscopic System (H.E.S.S.) telescope array, we developed a new multi-blob synchrotron self-Compton model.
  
  In the framework of this model, we present here the predictions for the very high energy emission of three active galactic nuclei with extended optical or X-ray jets which could be misaligned blazars, namely Cen\,A, PKS\,0521$-$36 and 3C\,273.
  
  
\end{abstract}
%
\section{Introduction}

The detection of M\,87 at very high energies (VHE; $> 100$GeV) by HEGRA (Aharonian et al. 2003), confirmed by H.E.S.S. (Aharonian et al. 2006), made it the first extragalactic object detected at the TeV range that is not a blazar. This discovery is probably the first insight toward further detections of objects already known to have an extended optical/X-ray jet with a moderate beaming towards us, but which are not genuine BL\,Lac objects.

Predictions of the VHE emission in the framework of a new synchrotron self-Compton (SSC) emission model of the jet for active galactic nuclei (AGNs) whose jet is not aligned with the line of sight, are presented here for 3 AGNs, Cen\,A, 3C\,273 and PKS\,0521$-$36.

In Sect.~\ref{sect:model}, we briefly present our SSC model, in Sect.~\ref{sect:3C273}, \ref{sect:CenA} and \ref{sect:PKS0521}, we present applications of this model to 3C\,273, Cen\,A and PKS\,0521$-$36 respectively, and conclusions are given in Sect.~\ref{sect:concl}.

\section{The multi-blob model}
\label{sect:model}

We developed a synchrotron self-Compton emission model for extragalactic object which jet is not aligned with the line of sight. The present work is based on the study of Katarzy{\'n}ski et al. (2001, 2003), who describe the radiative transfer of a single spherical relativistic blob of plasma moving along the jet axis. The blob is immersed in a uniform magnetic field. The emission from radio to optical/UV wavelengths is attributed to an inhomogeneous conical extended jet.

In our model, a cap, located in the large opening angle formation zone at the base of the jet, is filled with several blobs of plasma that radiate through synchrotron and inverse Compton scattering on the synchrotron photons (synchrotron self-Compton) over all frequencies from radio to VHE. This geometry allows to have some blobs that propagate close to the line of sight, even for sources which jet is misaligned, and thus are Doppler boosted, as in the case of blazars.
For more details on the multi-blob model, see Lenain (2007) and Lenain et al. (2007).

\section{3C\,273}
\label{sect:3C273}

\begin{figure}[h]
  \centering
  \includegraphics[angle=-90,width=9cm]{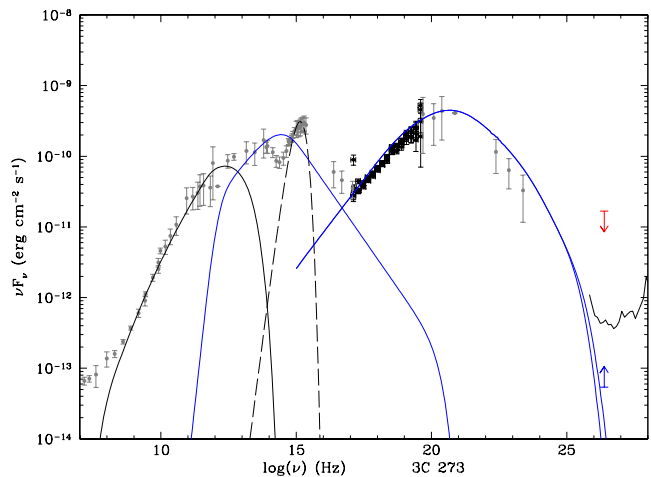}
  \caption{Spectral energy distribution of 3C\,273 with anticipated VHE flux.}
  \label{fig:1}
\end{figure}

3C\,273 ($z=0.158$) is the first quasar that was identified as a high-redshift object. The maximum acceptable viewing angle of its jet is about $15^\circ$ (Unwin et al. 1985). In Fig.~\ref{fig:1}, we show the spectral energy distribution (SED) with the anticipated VHE flux in blue as computed with the multi-blob SSC model. The upper limit in red was obtained at 3$\sigma$ by H.E.S.S. in 2005 (Aharonian et al. 2005), all the other data in gray are taken from T\"urler et al. (1999). The V-shaped curve at VHE shows the H.E.S.S. sensitivity limit for a detection at 5$\sigma$ in 50\,h of observation for a source at a mean zenith angle of $30^\circ$. The expected sensitivity of the next generation \v{C}erenkov Telescope Array (CTA) project of $\sim 0.1\%$\,Crab flux at 1\,TeV in 50\,h of observation is shown as a blue lower limit. The dashed line represents a simple blackbody model to illustrate the contribution of the big blue bump component in the UV. The parameters of the model are shown in Table~\ref{table:1}, where $\Gamma_b$ is the Lorentz factor common to each blob, $\theta$ is the viewing angle with respect to the jet axis, $R_\mathrm{cap}$ is the distance between the central engine and the cap of blobs, $r_g=G M_\mathrm{BH} / c^2$ is the scale length with $M_\mathrm{BH}$ the mass of the central black hole, $B$ is the intensity of the magnetic field, $r_b$ is the radius of an individual blob, $K_1$, $n_1$, $n_2$, $\gamma_\mathrm{min}$, $\gamma_\mathrm{br}$ and $\gamma_c$ are the parameters describing the electron energy distribution which is parametrized as a broken power-law.

As can be seen in Fig.~\ref{fig:1}, a strong detection of 3C\,273 by the current \v{C}erenkov facilities would be difficult to interpret within our nuclear multi-blob scenario. One solution would be to invoke different electron energy distribution among the blobs, which could result in a tail of the inverse Compton bump at VHE that could account for an eventual VHE detection. Another alternative would be the presence of an external inverse Compton component, which is then thought to be only weakly variable. Further observations with {\it GLAST} (10\,keV--300\,keV) and H.E.S.S.\,II, which will extend the spectral domain of H.E.S.S. down to $\sim 20$\,GeV with a better sensitivity, are crucial to distinguish between the different scenarii.

\section{Cen\,A}
\label{sect:CenA}

Cen\,A is the nearest radiogalaxy ($z=0.0018$). Although it is a nearby and well studied object, the value of the viewing angle of its jet is still controversial. Tingay et al. (1998) claim $\theta \sim 50^\circ$--$80^\circ$ for the parsec-scale jet, whereas Hardcastle et al. (2003) found $\theta \sim 15^\circ$ for the 100\,pc scale jet. We choose here to take an intermediate value of $\theta \sim 25^\circ$.

\begin{figure}[h]
  \centering
  \includegraphics[angle=-90,width=9cm]{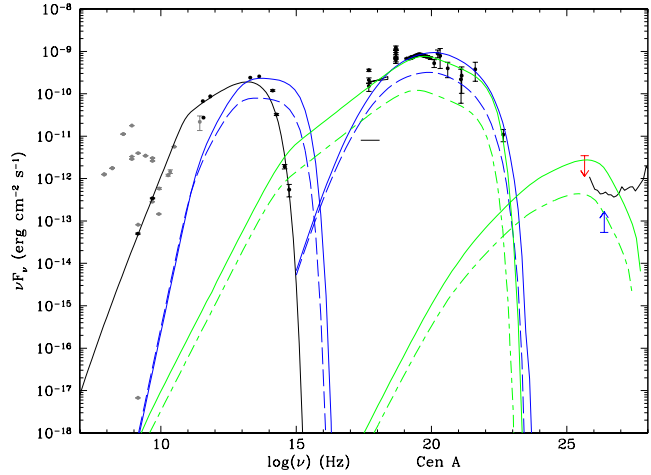}
  \caption{Spectral energy distribution of Cen\,A with anticipated VHE flux within the multi-blob SSC model.}
  \label{fig:2}
\end{figure}

Figure~\ref{fig:2} presents the SED of Cen\,A with anticipated VHE flux, where the data sample shown in black is almost the same as in Chiaberge et al. (2001), while the gray points are taken from the NED. We present in Fig.~\ref{fig:2} two solutions of the multi-blob model. The first one, presented in blue, assumes that the X-rays are generated by inverse Compton scattering. The second one, in green, assumes that the X-rays are of synchrotron origin. If the inverse Compton scattering produces emission in soft $\gamma$-rays, the SSC emission of the central region would definitely not provide a flux sufficient to be detected at VHE by current facilities. In that case, a detection at VHE could favor the presence of an external inverse Compton component, such as on the starlight photons of the host galaxy. On the other hand, if the soft $\gamma$-rays are of synchrotron origin, then we expect a detection of the core of Cen\,A at VHE by H.E.S.S. within 50\,h of observation.

\section{PKS\,0521$-$36}
\label{sect:PKS0521}

PKS\,0521$-$36 is a flat spectrum radio quasar (FSRQ, $z=0.055$) with an optical jet. No superluminal motion is seen in its jet, the beaming effect is thus much less important than in the case of 3C\,273. Indeed, the only constraint on the jet orientation comes from Pian et al. (1996) who find $\theta \simeq 30^\circ \pm 6^\circ$ from SSC models. It should also be noted that this object oscillates between a Seyfert-like and a BL\,Lac state (Ulrich 1981), making it difficult to interpret within a purely non-thermal scenario, especially since we are confronted to non-simultaneous data.

\begin{figure}[h]
  \centering
  \includegraphics[angle=-90,width=9cm]{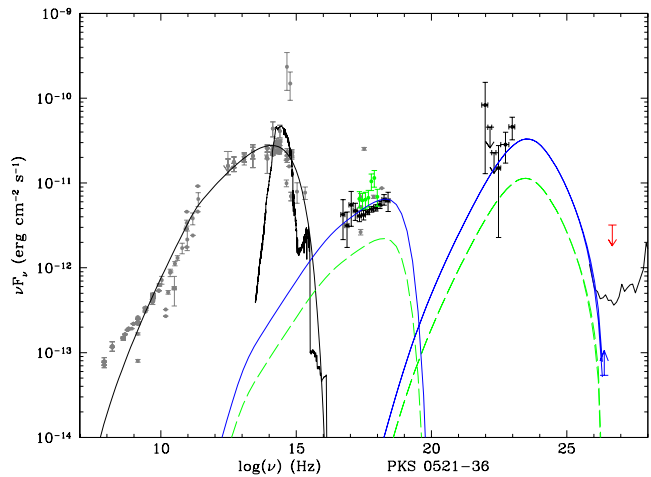}
  \caption{Spectral energy distribution of PKS\,0521$-$36 with the multi-blob SSC model.}
  \label{fig:3}
\end{figure}

Figure~\ref{fig:3} presents the SED of PKS\,0521$-$36 with anticipated VHE flux as computed with the multi-blob model. The black points are X-ray data taken by {\it Beppo}SAX in 1998 and by EGRET in 1994, while the gray points provided by the NED show the radio and optical data. The green points are more recent X-ray measurements by {\it Swift}/XRT in 2005. The black line in radio represents a model of synchrotron emission of an extended conical jet, while the black line in optical is a rough modeling of the host galaxy. The blue and green lines are two extreme geometric cases within the multi-blob model in which the line of sight is aligned with the velocity vector of a blob (blue line) or passes exactly through the gap between three blobs, thus leading to a weaker beaming effect (green line).

If the X-rays were due to inverse Compton scattering, it would be on photons from the radio/optical contribution which is not variable, coming from an extended part of the jet. Since the X-rays are highly variable in this object, they are likely due to the synchrotron process. In this case, PKS\,0521$-$36 should be marginally detectable by current \v{C}erenkov such as H.E.S.S. and easily detectable by H.E.S.S.\,II and CTA.

\begin{table}[h]
\caption{Parameters used in Figs.~\ref{fig:1}, \ref{fig:2} and \ref{fig:3}}
\label{table:1}
\centering
\begin{tabular}{c c c c c}
\hline\hline
&  3C\,273 & Cen\,A (\it blue) & Cen\,A (\it green) & PKS\,0521$-$36\\
\hline
$\Gamma_b$ & 7.4 & 8.14 & 20.0 & 1.5\\
$\theta$ & 15$^\circ$ & 25$^\circ$ & 25$^\circ$ & 25$^\circ$\\
$R_\mathrm{cap}$ [$r_g$] & 100.0 & 100.0 & 100.0 & 100.0\\
$B$ [G] & 3.0 & 2.0 & 10.0 & 1.0\\
$r_b$ [cm] & $2.0 \times 10^{15}$ & $1.0 \times 10^{14}$ & $8.0 \times 10^{13}$ & $9.0 \times 10^{14}$\\
$K_1$ [cm$^{-3}$] & $1.8 \times 10^6$ & $9.0 \times 10^7$ & $4.0 \times 10^4$ & $3.0 \times 10^6$\\
$n_1$ & 2.0 & 2.0 & 2.0 & 2.0\\
$n_2$ & 4.1 & 3.0 & 3.5 & 2.5\\
$\gamma_\mathrm{min}$ & $1$ & $3.0 \times 10^2$ & $10^3$ & $10^3$\\
$\gamma_\mathrm{br}$ & $1.6 \times 10^3$ & $5.0 \times 10^2$ & $3.5 \times 10^5$ & $5.0 \times 10^4$\\
$\gamma_c$ & $10^6$ & $4.0 \times 10^3$ & $6.0 \times 10^6$ & $10^6$\\
\hline
\end{tabular}
\end{table}

\section{Conclusions}
\label{sect:concl}

We have presented a SSC model to anticipate the VHE flux emission of 3 misaligned AGNs with extended jets. This model accounts in a simple way for a differential Doppler boosting by modeling the radiative transfer in several blobs of plasma located in the broadened zone at the base of the jet. This scenario provides the possibility to extend standard leptonic models of TeV blazars to other types of AGNs which jet is misaligned. However we can not exclude other leptonic or hadronic models. More observations are needed to constrain the emission models and to distinguish between hadronic and leptonic scenarii. This will be hopefully feasible with the next generation of $\gamma$-ray facilities, such as {\it GLAST}, H.E.S.S.\,II and CTA.

\begin{acknowledgements}
  The author would like to thank Dr.~C.~Boisson and Dr.~H.~Sol for their invaluable help, and Dr.~A.~Djannati-Ata{\"i}, Dr.~S.~Pita and Dr.~A.~Zech for useful discussions.
  
  This research has made use of the NASA/IPAC Extragalactic Database (NED) which is operated by the Jet Propulsion Laboratory, California Institute of Technology, under contract with the National Aeronautics and Space Administration.
\end{acknowledgements} 



\end{document}